\documentclass[aps,prd,twocolumn,preprintnumbers,longbibliography,10pt]{revtex4-2}
\usepackage[colorlinks=true, linkcolor=blue, citecolor=blue, urlcolor=blue]{hyperref}
\usepackage{bm,latexsym,amssymb,amsmath,mathrsfs}
\usepackage{graphicx}
\usepackage{color}
\usepackage{times,setspace}
\usepackage{physics}
\usepackage{mathtools}

\begin{document}
\preprint{YITP-24-115}
\title{Quantum circuit for \texorpdfstring{$\mathbb{Z}_3$}{Z3} lattice gauge theory at nonzero baryon density}

\author{Yoshimasa Hidaka}
\affiliation{Yukawa Institute for Theoretical Physics, Kyoto University, Kyoto 606-8502, Japan}
\affiliation{Interdisciplinary Theoretical and Mathematical Sciences Program (iTHEMS), RIKEN, Wako, Saitama 351-0198, Japan}

\author{Arata Yamamoto}
\affiliation{Department of Physics, The University of Tokyo, Tokyo 113-0033, Japan}

\begin{abstract}
$\mathbb{Z}_3$ lattice gauge theory is the simplest discrete gauge theory with three-quark bound states, i.e., baryons.
Since it has a finite-dimensional Hilbert space, it can be used for testing quantum simulation of lattice gauge theory at nonzero baryon density.
We discuss global and local gauge symmetries and their importance in quantum simulation.
We perform quantum emulator calculation and demonstrate how to study the ground state property of baryonic matter.
\end{abstract}

\maketitle

\section{Introduction}

Quantum computing will be a technological innovation for lattice gauge theory~\cite{Banuls:2019bmf,Bauer:2022hpo,DiMeglio:2023nsa}.
In the future, it might help us solve long-standing mysteries in particle and nuclear physics; e.g., properties of dense baryonic matter in quantum chromodynamics (QCD) \cite{Yamamoto:2021vxp,Tomiya:2022chr,Davoudi:2022uzo,Schuster:2023klj}.
Although quantum simulation of QCD is too challenging for current technology, we should take a step closer to the ultimate goal.
The $SU(3)$ gauge group is theoretically complicated and practically hard due to its infinite-dimensional Hilbert space \cite{Ciavarella:2021nmj,Ciavarella:2021lel,Kadam:2022ipf,Hayata:2023puo,Hayata:2023bgh,Ciavarella:2024fzw,Bergner:2024qjl}.
Discrete gauge groups are good starting points for near-term quantum computing; for instance, $\mathbb{Z}_N$ \cite{Zohar:2016wmo,Ercolessi:2017jbi,Lamm:2019bik,Magnifico:2019kyj,Yamamoto:2020eqi,Gustafson:2020yfe,Homeier:2022mkg,Irmejs:2022gwv,Sukeno:2022pmx,Hayata:2023skf}, $D_N$ \cite{Lamm:2019bik,Alam:2021uuq,Ballini:2023ljs}, $\mathbb{BO}$ \cite{Gustafson:2023kvd}, $\mathbb{BT}$ \cite{Gustafson:2022xdt,Lamm:2024jnl}, $\mathbb{BI}$ \cite{Lamm:2024jnl}, and $\Sigma(3N)$ \cite{Alexandru:2019nsa,Gustafson:2024kym}.

$\mathbb{Z}_2$ lattice gauge theory is most often used for designing quantum simulations because one $\mathbb{Z}_2$ gauge link is encoded using one qubit.
Because of its simplicity, simulations of $\mathbb{Z}_2$ lattice gauge theory perform well on noisy intermediate-scale quantum (NISQ) devices \cite{Mildenberger:2022jqr,Pardo:2022hrp,Charles:2023zbl,Hayata:2024smx}.
In $\mathbb{Z}_2$ lattice gauge theory, however, a baryon is a bound state of two quarks.
Since such a bosonic baryon does not form a Fermi surface, the behavior of its dense matter is completely different from the fermionic baryonic matter in the real world.
$\mathbb{Z}_2$ lattice gauge theory cannot be used for studying nonzero baryon density.
This discrepancy is resolved by considering $\mathbb{Z}_3$ lattice gauge theory \cite{Hidaka:2024drb,Florio:2023kel}.
In $\mathbb{Z}_3$ lattice gauge theory, three quarks are charge neutral and from a fermionic bound state due to confining force.
$\mathbb{Z}_3$ lattice gauge theory can be used for a toy model for QCD at nonzero baryon density \cite{Hidaka:2024drb}.
The one-dimensional $\mathbb{Z}_3$ lattice gauge theory will be a feasible setup for benchmarking quantum simulations on NISQ devices.
($\mathbb{Z}_3$ lattice gauge theory is also valuable for a minimal model of the sign problem in the Monte Carlo sampling~\cite{Gattringer:2012jt,Akiyama:2023hvt}.)

In this paper, we design a quantum simulation of $\mathbb{Z}_3$ lattice gauge theory at nonzero baryon density.
We begin with the Hamiltonian and symmetries of the one-dimensional $\mathbb{Z}_3$ lattice gauge theory in Sec.~\ref{secZ3}.
To understand the importance of gauge symmetry, we study the energy spectrum at nonzero baryon density in Sec.~\ref{secenergy}.
We construct gauge invariant protocols for a quantum circuit in Sec.~\ref{secpro} and test ground-state calculations using the Qiskit simulator in Sec.~\ref{sectest}.
Finally, we provide technical comments in Sec.~\ref{seccom}.
All equations are written in the lattice unit and all quantities are dimensionless throughout the paper.

\section{\texorpdfstring{$\mathbb{Z}_3$}{Z3} lattice gauge theory in \texorpdfstring{$(1+1)$}{(1+1)} dimensions}
\label{secZ3}

The theory is formulated on a one-dimensional lattice, $x=1,2,\cdots, L$.
In $\mathbb{Z}_3$ lattice gauge theory, the gauge link operator $U(x)$ and the conjugate operator $\Pi(x)$ satisfy the relation
\begin{equation}
\Pi(x) U(x) = \omega U(x) \Pi(x)
\end{equation}
with $\omega=\exp (2\pi i/3)$ \cite{Horn:1979fy}.
The Dirac fermion operators $\psi_q(x)$ satisfy the anticommutation relation
\begin{equation}
 \{ \psi_{q\alpha}(x), \psi_{q'\beta}^\dagger(x) \} = \delta_{qq'}\delta_{\alpha\beta},
\end{equation}
where the spinor indices $\alpha$ and $\beta$ run from $1$ to $2$.
The Dirac fermion is three-flavor, $q=u,d,s$, such that the lightest baryon is s-wave \cite{Florio:2023kel}.
Adopting the Wilson fermion formalism, we consider the Hamiltonian
\begin{equation}
\begin{split}
\label{eqH}
 H =& \sum_{x} g^2 \left[ 1 - \frac{1}{2} \{ \Pi (x) + \Pi^\dagger (x) \} \right] \\
&+ \sum_{x,q} \Big\{ (1+m)  \psi_q^\dagger(x) \gamma^0 \psi_q(x) \\
&-\frac{1}{2} \psi_q^\dagger(x) \gamma^0 (1-i \gamma^1) U(x) \psi_q(x+1) \\
&-\frac{1}{2} \psi_q^\dagger(x+1) \gamma^0 (1+i \gamma^1) U^\dagger(x) \psi_q(x) \Big\}
\end{split}
\end{equation}
with the gauge parameter $g$ and the quark mass $m$.
Here, $\gamma^\mu$ are the Gamma matrices, which satisfy $(\gamma^0)^2=-(\gamma^1)^2=1$, and $\gamma^0\gamma^1=-\gamma^1\gamma^0$.

The theory has global $U(3)$ flavor symmetry,
\begin{equation}
\psi_q(x)\to W_\theta\psi_q(x),\quad \psi_q^\dag(x)\to W_\theta^\dag\psi_q^\dag(x),
\end{equation}
and local $\mathbb{Z}_3$ gauge symmetry,
\begin{align}
\psi_q(x)&\to \omega^{n_x}\psi_q(x),\quad \psi_q^\dag(x)\to \omega^{-n_x}\psi_q^\dag(x),\\
U(x)& \to \omega^{n_x-n_{x+1}}U(x),
\end{align}
where $W_\theta$ is a $U(3)$ flavor matrix, and $n_x$ is an integer depending on the space coordinate.
The flavor symmetry implies that the eigenstates of Hamiltonian are classified by the irreducible representation of $U(3)$ group.
In particular, the diagonal part, $U(1)\times U(1)\times U(1)$, protects quark numbers.
The quark number operator of flavor $q$ is defined by
\begin{equation}
 Q_q =  \sum_x \{ \psi^\dagger_q(x) \psi_q(x) -1 \}.
\end{equation}
Because of the commutation relation $[H,Q_q]=0$, the energy eigenstates have fixed quark numbers,
\begin{align}
 H|\Psi\rangle &= E|\Psi\rangle,
\\
 Q_q|\Psi\rangle &=N_q|\Psi\rangle,
\end{align}
 for all $q$.
All the quark numbers $(N_u,N_d,N_s)$ are individually conserved and the baryon number
\begin{equation}
 B= \frac{1}{3} \sum_q N_q= \frac{1}{3} (N_u+N_d+N_s)
\end{equation}
is conserved.
Note that these quantum numbers do not fully determine the irreducible representation. In the case of $U(3)$, it is also necessary to specify the values of the second and third Casimir invariants in order to obtain an irreducible representation.

The gauge symmetry is defined by the Gauss law operator
\begin{equation}
G(x) = \Pi(x)\Pi^\dag(x-1) \omega^{-\rho(x)}
\label{eqG}
\end{equation}
with
\begin{equation}
\rho(x) = \sum_q\{\psi^\dag_q(x)\psi_q(x)-1\}.
\end{equation}
Physical, i.e., gauge invariant, eigenstates satisfy the constraint
\begin{equation}
G(x)|\Psi\rangle =|\Psi\rangle
\end{equation}
for all $x$ but unphysical eigenstates do not satisfy it.
The Gauss law operator commutes with the Hamiltonian, $[H,G(x)]=0$.
The operator $G(x)$ corresponds to the exponential of the Gauss law in continuous gauge theory.
$\Pi(x)$ is the exponential of an electric field and $\frac{2\pi}{3} \rho(x)$ is a local charge.
The eigenvalues of the electric field and the local charge are defined in mod $2\pi$ and take only three values, $0$, $\frac{2\pi}{3}$, and $\frac{4\pi}{3}$.
This is nothing but finite-dimensional nature of discrete gauge theory.
(The electric field has infinitely many eigenvalues in continuous gauge theory, and these operators are noncommutative in nonabelian gauge theory.)

\section{Energy spectrum}
\label{secenergy}

\begin{figure}[ht!]
\begin{center}
 \includegraphics[width=0.48\textwidth]{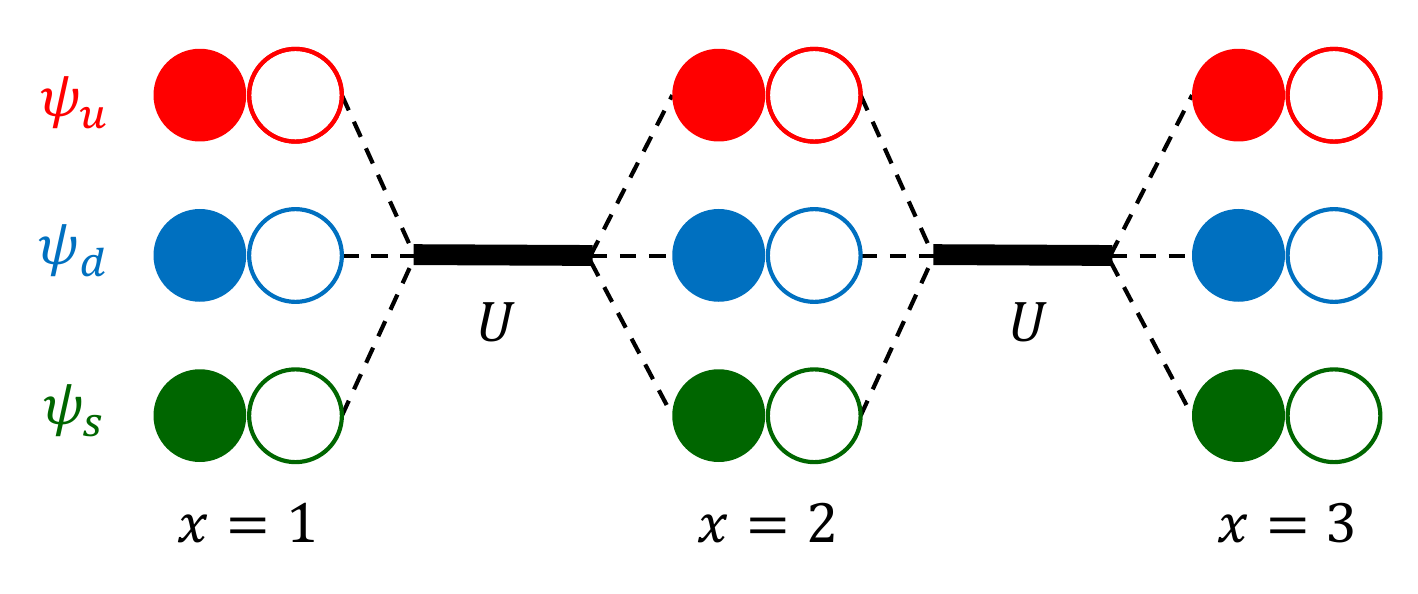}
\end{center}
\caption{
\label{figlattice}
Schematic figure of the one-dimensional lattice with open boundary conditions.
The filled and unfilled circles represent upper and lower spinor components, respectively.
}
\end{figure}

Before discussing quantum computing, we study this theory by the brute-force matrix calculation on a classical computer.
For concreteness, we specify lattice geometry as shown in Fig.~\ref{figlattice}.
The lattice is one-dimensional with open boundaries.
The number of sites is $L=3$, and the number of links is $L-1=2$.
The maximum baryon number is $B=3$ and the minimum baryon number is $B=-3$.
The total dimension of the Hilbert space is $D = 2^{6L} \times 3^{L-1}=2,359,296$.
When system size is small, we can write down a fixed-quark-number basis, i.e., the basis that labels which lattice sites are occupied or unoccupied by quarks.
We use the fixed-quark-number basis and decompose into fixed-quark-number sectors for classical matrix calculation in this section.
This reduces the fermion matrix size from $2^{18}$ to $\binom{6}{3}^3=8,000$ in the vacuum sector $(N_u=N_d=N_s=0)$ etc.
We fix the gauge coupling constant at $g^2=1$ and the quark mass at $m=0.1$.

\begin{figure}[ht]
\begin{center}
 \includegraphics[width=0.48\textwidth]{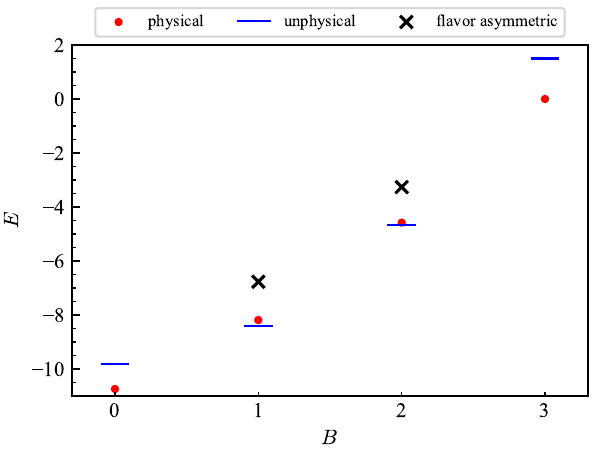}
\end{center}
\caption{
\label{figdiag}
Lowest eigenenergies $E$ of physical states (red dots) and unphysical states (blue bars) in the flavor symmetric sectors $N_u=N_d=N_s$.
The lowest eigenenergies of physical states in flavor asymmetric sectors are also shown (black crosses).
}
\end{figure}

The flavor symmetric sectors, $N_u=N_d=N_s$, are the most important.
Figure~\ref{figdiag} shows the ground state energies obtained by the exact matrix diagonalization.
In the vacuum sector $B=0$, the physical ground state has a lower energy than the unphysical ground state.
This is consistent with our intuition; the vacuum is gauge invariant.
For nonzero densities, $B=1$ and $2$, the physical ground states have slightly higher energy than the unphysical ground states.
This can be understood in strong coupling picture.
In the strong coupling limit $g \to \infty$, electric fields must vanish, $\Pi(x)|\Psi\rangle=|\Psi\rangle$, to minimize electric field energy.
As for physical states, the Gauss law constraint forces up, down, and strange quarks to form a point-like baryon.
As for unphysical states, since quark positions are not constrained by electric fields, delocalized quarks are possible and energetically favored.
The delocalized quarks have lower fermion energy than the point-like baryon of the physical state.

This observation is crucial for the numerical simulation of the ground state at nonzero baryon density.
The lowest energy state in a fixed-quark-number sector is unphysical.
The physical ground state cannot be obtained by naive energy minimization in the total Hilbert space, but can be obtained only by restricting to the gauge-invariant subspace.
For example, in adiabatic simulations, the ground state is obtained by applying evolution operators to an arbitrary initial state.
The evolution operators must be gauge invariant and the initial state must be physical; otherwise the resulting state will be unphysical.

The ground state energies in other flavor sectors are also shown in Fig.~\ref{figdiag}.
Flavor asymmetric sectors are energetically disfavored due to the Pauli exclusion principle.
For example, the ground state energy of $(N_u,N_d,N_s)=(2,1,0)$ is higher than that of $(1,1,1)$, as shown in the figure.
That of $(3,0,0)$ is even higher.
The $B=1$ sector consists of three quarks, whose representation is decomposed as $\bm{3}\otimes \bm{3}\otimes \bm{3}= \bm{10}\oplus\bm{8}\oplus\bm{8}\oplus\bm{1}$.
For a fixed quark number, $(3,0,0)$ contains only $\bm{10}$ representation, while $(2,1,0)$ contains $\bm{10}$ and $\bm{8}$ representations.
By contrast, the flavor symmetric sector $(1,1,1)$ contains all possible representations: 
$\bm{10}$, $\bm{8}$, and $\bm{1}$.
These observations imply that the ground states for $(N_u,N_d,N_s)=(3,0,0)$, $(2,1,0)$, and $(1,1,1)$ correspond to $\bm{10}$, $\bm{8}$, and $\bm{1}$ representations, respectively.
We focus only on the flavor symmetric sectors $N_u=N_d=N_s$ in the following sections.

\section{Gauge invariant protocols}
\label{secpro}

Preserving global and local gauge symmetries is important for designing quantum circuits.
If the symmetries are violated, a computed state might be a mixture of different quark numbers or unphysical states.
In many algorithms of quantum simulation, such as time evolution and ground-state calculation, we need the exponential operators of individual terms in a Hamiltonian.
We here construct gauge invariant circuits for the exponential operators in the $\mathbb{Z}_3$ lattice gauge theory.

Each gauge link is three-dimensional,
and the state vector is represented by two qubits $|g_0\rangle \otimes |g_1\rangle$ as
\begin{equation}
|00\rangle =
\begin{pmatrix}
   1 \\
   0 \\
   0 
\end{pmatrix}
, \quad |01\rangle =
\begin{pmatrix}
   0 \\
   1 \\
   0 
\end{pmatrix}
, \quad |11\rangle =
\begin{pmatrix}
   0 \\
   0 \\
   1
\end{pmatrix}.
\end{equation}
For the gauge operators, we choose the $U$-diagonal basis
\begin{equation}
\label{eqbasis}
U =
    \begin{pmatrix}
   1 & 0 & 0 \\
   0 & \omega & 0 \\
   0 & 0 & \omega^2 \\
    \end{pmatrix}
, \quad \Pi =
    \begin{pmatrix}
   0 & 1 & 0 \\
   0 & 0 & 1 \\
   1 & 0 & 0 \\
    \end{pmatrix} .
\end{equation}
The argument $x$ is omitted here.
The gauge part of the Hamiltonian \eqref{eqH} can be decomposed into three flip matrices as
\begin{equation}
  \Pi + \Pi^\dagger = X_a + X_b + X_c
\end{equation}
with
\begin{equation}
X_a =
\begin{pmatrix}
   0 & 1 & 0 \\
   1 & 0 & 0 \\
   0 & 0 & 0
\end{pmatrix}
, \ X_b =
\begin{pmatrix}
   0 & 0 & 0 \\
   0 & 0 & 1 \\
   0 & 1 & 0
\end{pmatrix}
, \ X_c =
\begin{pmatrix}
   0 & 0 & 1 \\
   0 & 0 & 0 \\
   1 & 0 & 0
\end{pmatrix}.
\label{eq:Xabc}
\end{equation}

The Hamiltonian \eqref{eqH} consists of three parts: the gauge part, the fermion mass part, and the fermion hopping part.
Each part is individually global and local gauge invariant unless further decomposed into gauge variant components.
The exponential operator of the gauge part is rewritten as
\begin{align}
\label{eqeg}
e^{i\theta(\Pi+\Pi^{\dag})}= V \Lambda(\theta) V^\dag,
\end{align}
where
\begin{align}
V&=\frac{1}{\sqrt{3}}
\begin{pmatrix}
1 & 1 & i\\
1 & \omega & i\omega^2\\
1 & \omega^2 & i\omega
\end{pmatrix}
=e^{-i\frac{\pi}{4}Y_b} e^{-i\varphi Y_a} e^{-i\frac{3\pi}{4} X_b},
\\
\Lambda&= 
\begin{pmatrix}
e^{2i\theta} & 0 & 0\\
0 & e^{-i\theta} & 0\\
0 & 0 & e^{-i\theta}
\end{pmatrix}
= e^{i\theta Z_a} e^{i\theta Z_c},
\end{align}
with $\varphi=\cos^{-1}(1/\sqrt{3})$.
The matrices $Y_a$, $Y_b$, $Z_a$, and $Z_c$ are similarly defined as in Eq.~\eqref{eq:Xabc}.
The circuits are constructed as
\begin{align}
 e^{i\theta X_b} &= \parbox[c]{150pt}{\includegraphics[scale=0.8]{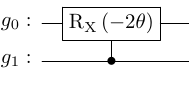}}
\\
 e^{i\theta Y_a} &= \parbox[c]{150pt}{\includegraphics[scale=0.8]{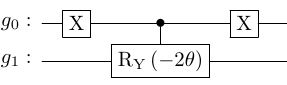}}
\\
 e^{i\theta Y_b} &= \parbox[c]{150pt}{\includegraphics[scale=0.8]{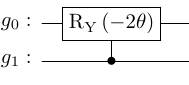}}
 \\
 e^{i\theta Z_a} &= \parbox[c]{150pt}{\includegraphics[scale=0.8]{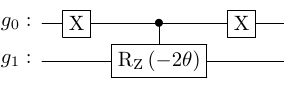}}
\\
 e^{i\theta Z_c} &= \parbox[c]{150pt}{\includegraphics[scale=0.8]{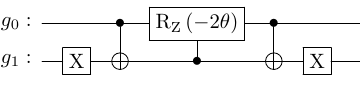}}
.
\end{align}
For the fermion part, taking $\gamma^0=\sigma_1$ and $i\gamma^1=\sigma_3$, we can rewrite the mass term as
\begin{equation}
\psi_q^\dagger(x) \gamma^0 \psi_q(x) = \psi^\dag_{2q}(x) \psi_{1q}(x) + \psi^\dag_{1q}(x) \psi_{2q}(x),
\end{equation}
and the hopping term as
\begin{equation}
\begin{split}
& \frac12 \{ \psi_q^\dagger(x) \gamma^0 (1-i \gamma^1) U(x) \psi_q(x+1) \\
& \quad + \psi_q^\dagger(x+1) \gamma^0 (1+i \gamma^1) U^\dagger(x) \psi_q(x) \} \\
&= \psi^\dag_{1q}(x)U(x)\psi_{2q}(x+1) + \psi^\dag_{2q}(x+1)U^\dag(x)\psi_{1q}(x) .
\end{split}
\end{equation}
The fermions are arranged on a one-dimensional chain and qubitized by the standard Jordan-Wigner transformation.
Both the mass and hopping terms act on two neighboring fermions.
The exponential operators acting on $|f_0\rangle \otimes |f_1\rangle$ are implemented as
\begin{align}
\label{eqem}
&e^{i\theta ( \psi^\dag_{1}\psi_{0} + \psi^\dag_{0}\psi_{1} ) }=\parbox[c]{50pt}{\includegraphics[scale=0.6]{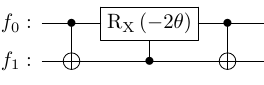}}
\\
\label{eqeu}
\begin{split}
&e^{i\theta ( \psi^\dag_{0}U\psi_{1} + \psi^\dag_{1}U^\dag\psi_{0} ) }\\
&=\parbox[c]{200pt}{\includegraphics[scale=0.6]{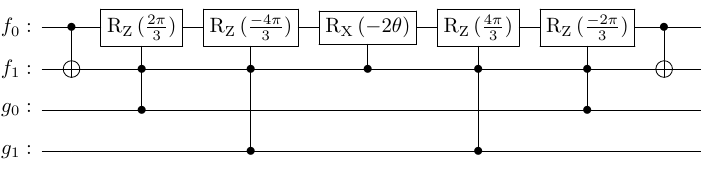}}
\end{split}
\end{align}
for the $U$-diagonal basis \eqref{eqbasis}.
The exponential operators \eqref{eqeg}, \eqref{eqem}, and \eqref{eqeu} commute with the Gauss law operator $G(x)$ and the quark number operators $Q_q$.
The circuit constructed from them conserves the Gauss law and the quark numbers.

While we use the $U$-diagonal basis in this paper, we can also work with the $\Pi$-diagonal basis.
The matrix forms of $U$ and $\Pi$ flip in Eq.~(15), and the diagonalizing unitary matrix appears not in the gauge part but in the fermion hopping part.
Although optimal basis choice depends on the details of simulation setup, there is no significant difference in most cases.

\section{Emulator test}
\label{sectest}

We tested the quantum simulation by utilizing the noiseless Qiskit simulator.
The lattice geometry, the gauge parameter, and the quark mass are the same as those in Sec.~\ref{secenergy}.
Although gauge fields can be eliminated in this geometry, we keep and treat the gauge fields to demonstrate the gauge invariant protocols.
The simulation requires 22 qubits: 18 qubits for fermions and 4 qubits for gauge fields.
We simulated $B=0$, $1$, and $2$ for the flavor symmetric sectors.
The fully occupied state $B=3$ can be analytically solvable and its energy is zero.

We computed the ground states at nonzero baryon density by the quantum adiabatic algorithm \cite{Albash:2016zte}.
The ground state $|\Psi\rangle$ is obtained by the adiabatic evolution
\begin{align}
\label{eqevol}
 |\Psi\rangle &= e^{-i \int_0^S ds h (s)} |\Phi\rangle, \\
 h(s) &= \left( 1-\frac{s}{S} \right) H_0 + \frac{s}{S} H,
\end{align}
where and $|\Phi\rangle$ is the ground state of the initial Hamiltonian
\begin{equation}
\begin{split}
H_0 =&  \sum_{x=1}^2 g^2 \left[ 1 - \frac{1}{2} \{ \Pi (x) + \Pi^\dagger (x) \} \right] \\
&+ \sum_{x=1}^3 \sum_q \{1+m+v(x)\}  \psi_q^\dagger(x) \gamma^0 \psi_q(x)
\end{split}
\end{equation}
with
\begin{equation}
 v(x) =
\begin{cases}
0 & (B = 0)\\
-m\delta_{x2} & (B = 1)\\
+m\delta_{x2} & (B = 2)
\end{cases}.
\end{equation}
The initial state is given by zero electric fields, $\Pi(x)|\Phi\rangle=|\Phi\rangle$, and the flavor-singlet three quarks localized at $x=2$ ($B=1$) and at $x=1$ and $3$ ($B=2$), which can be prepared by a shallow circuit.
This satisfies the eigenvalue equation $Q_q|\Phi\rangle =N_q|\Phi\rangle$ and the Gauss law constraint $G(x)|\Phi\rangle =|\Phi\rangle$.
The evolution operator is approximated as $\exp \{ -i \int_0^S ds h (s)\} \simeq \exp\{-i \delta h (S)\} \cdots \exp\{-i \delta h (\delta)\} $ with step size $\delta=0.5$, and each evolution operator $\exp\{-i \delta h (s)\}$ is decomposed into the gauge part \eqref{eqeg}, the fermion mass part \eqref{eqem}, and the fermion hopping part \eqref{eqeu} by the second-order Lie-Suzuki-Trotter formula.
As discussed above, all the decomposed operators commute with the quark number operators and the Gauss law operator.
The obtained ground state has the same quark numbers as the initial state, $Q_q|\Psi\rangle =N_q|\Psi\rangle$, and satisfies the Gauss law constraint $G(x)|\Psi\rangle =|\Psi\rangle$.

\begin{figure}[ht]
\begin{center}
 \includegraphics[width=0.48\textwidth]{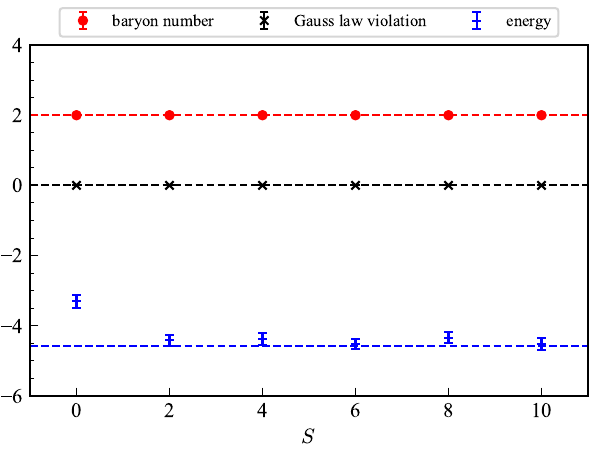}
\end{center}
\caption{
\label{figaqc}
Adiabatic evolution of the baryon number $B$, the Gauss law violation $C$, and the energy $E$ as functions of the evolution length $S$.
The error bars represent statistical uncertainties,
and the broken lines show the exact values.
}
\end{figure}

\begin{figure}[ht]
\begin{center}
 \includegraphics[width=0.48\textwidth]{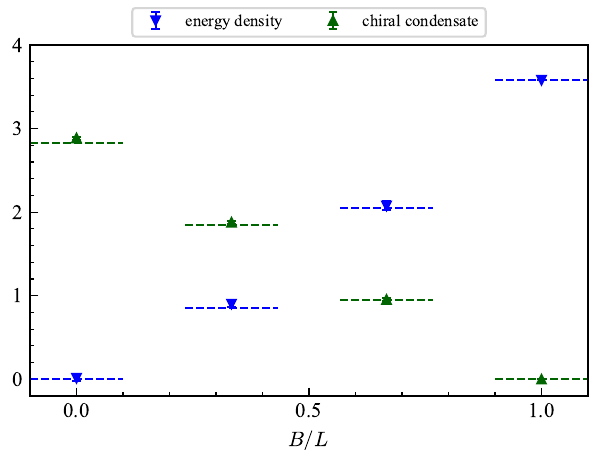}
\end{center}
\caption{
\label{figECB}
Energy density $\epsilon$ and chiral condensate $\Sigma$ as functions of baryon number density $B/L$.
The simulation data at $S=10$ are used.
The error bars represent statistical uncertainties,
and the broken lines show the exact values.
}
\end{figure}

Let us first check the gauge invariance along the evolution.
Figure \ref{figaqc} shows the evolution of observables: the baryon number $B=\frac{1}{3} \langle Q_u+Q_d+Q_s \rangle$, the Gauss law violation
\begin{equation}
  C=\sum_x \langle | G(x)-1 |^2  \rangle,
\end{equation}
and the energy $E=\langle H \rangle$.
The baryon number is conserved and the Gauss law violation is zero during the evolution.
The energy sufficiently converges to the exact ground-state energy at large $S$.
Therefore, the evolution is gauge invariant and the physical ground state is successfully obtained for a fixed baryon number.
The density dependence of physical observables can be studied from the obtained ground states, as shown in Fig.~\ref{figECB}.
The energy density is defined by
\begin{equation}
 \epsilon = \frac{1}{L} (E - E_{\rm vac}) ,
\end{equation}
where the exact vacuum energy $E_{\rm vac}$ is subtracted for visibility, and  the chiral condensate is defined by
\begin{equation}
 \Sigma = - \frac{1}{L} \sum_{x,q} \langle \psi_q^\dagger(x) \gamma^0 \psi_q(x) \rangle .
\end{equation}
The simulation reproduces the exact values for all the baryon numbers.
Currently, we are limited to small volumes due to the constraints of classical computing resources.
If quantum computers are available in the future, the volume can be taken larger and the thermodynamics limit can be extrapolated.

\begin{figure}[ht]
\begin{center}
 \includegraphics[width=0.48\textwidth]{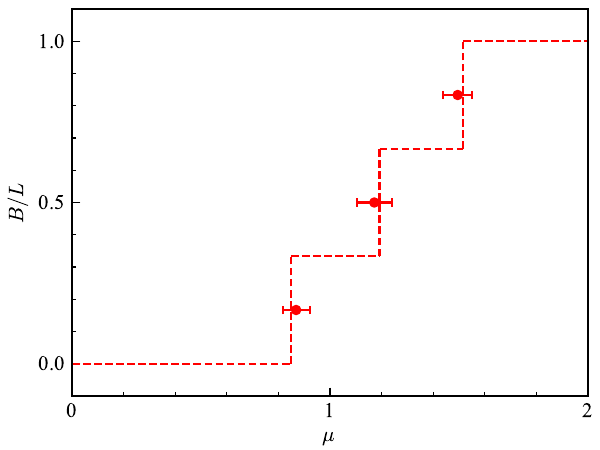}
\end{center}
\caption{
\label{figBmu}
Transition points of baryon number density $B/L$.
The simulation data at $S=10$ are used.
The error bars represent statistical uncertainties.
The broken line shows the exact functional form.
}
\end{figure}

It is also possible to convert the density dependence to the dependence on a quark chemical potential $\mu$.
At zero temperature, the relation between $B$ and $\mu$ is determined by minimizing
\begin{equation}
 \Omega(B,\mu) = E - \mu(N_u+N_d+N_s).
\end{equation}
The second term is a constant without statistical error in the fixed-quark-number simulation.
The ground state for a given $\mu$ has the smallest $\Omega(B,\mu)$ among all $B$.
A transition in $B$ happens at $\mu$ where $\Omega(B,\mu)$ and $\Omega(B+1,\mu)$ intersect.
Figure \ref{figBmu} shows the transition points obtained from the simulation data.
The plot is a sum of step functions in a finite box.
In the thermodynamic limit, it will become a frequently-seen plot of a nonzero-density phase transition.

\section{Comments}
\label{seccom}

The above computation is based on the gauge invariant circuit of the evolution equation \eqref{eqevol}.
Although the simulation is, in principle, possible even if the gauge invariance is lost, it will be less efficient.
If the global gauge symmetry is lost, the evolved state is not an eigenstate of the quark number operators.
One must compute the lowest eigenstate of $H-\mu (Q_u+Q_d+Q_s)$, instead of $H$, for many different values of $\mu$.
The quantum adiabatic algorithm assumes a nonzero gap $\Delta$ between the ground state and the first excited state and its convergence speed is proportional to $\Delta^2$ \cite{Albash:2016zte}.
The computational cost will be very large around the transition point of $\mu$, where the gap between $B$ and $B+1$ disappears.
The violation of the local gauge symmetry is more harmful.
The Gauss-law-violated evolution provides the lowest energy state in the total Hilbert space, including both physical and unphysical subspaces.
The obtained state is physical only when the unphysical ground state has higher energy than the physical ground state.
This condition is satisfied in the vacuum but not satisfied for nonzero density states.
The evolution must be modified with a projection operator onto the physical subspace.
This will cause extra computational costs.

Although our simulation is the classically emulated one, the same simulation will be possible on real qubit devices.
The required number of qubits is $6L+2(L-1)=22$.
To our knowledge, the optimal number of CNOT gates is $3(L-1)$ for Eq.~\eqref{eqeg}, $2L$ for Eq.~\eqref{eqem}, $19(L-1)$ for Eq.~\eqref{eqeu}, and thus $29L-25$ for each step of the evolution operator decomposed by the second-order formula.
As shown in Fig.~\ref{figaqc}, the evolution converges within a few steps in the case of $L=3$.
This is simulable even on the present NISQ devices by mitigating error \cite{Mildenberger:2022jqr,Pardo:2022hrp,Charles:2023zbl,Hayata:2024smx}.
For extrapolation to the the thermodynamic limit, we will need $L=O(10)$ and more steps, thus need a few hundred qubits and better gate fidelity.
This will be achievable in the near future and a nonzero-density phase transition will be found in the one-dimensional model.
In the distant future when a large amount of resources are available, we would like to extend the model to three spatial dimensions and study the toy model of dense QCD on quantum computers \cite{Hidaka:2024drb}.

\begin{acknowledgments}
The authors thank Yuya Tanizaki for fruitful discussions.
This work was supported by JSPS KAKENHI Grant No.~21H01084, 24H00975 (Y.~H.), and 19K03841 (A.~Y.), and by JST, CREST Grant No.~JPMJCR24I3, Japan.
The work of Y.~H.~was also partially supported by Center for Gravitational Physics and Quantum Information (CGPQI) at Yukawa Institute for Theoretical Physics.
The classical matrix diagonalization was performed on SQUID at the Cybermedia Center, Osaka University.
\end{acknowledgments}

\bibliographystyle{apsrev4-2}
\bibliography{paper}

\end{document}